\documentclass[pra, aps, showpacs, superscriptaddress, twocolumn]{revtex4-2}
\usepackage{graphicx,amsmath,amssymb,color}
\usepackage{physics}
\usepackage{graphicx}
\usepackage{epstopdf}
\usepackage{float}
\usepackage{xcolor}
\usepackage{soul}
\usepackage[colorlinks, citecolor={blue!50!black}, urlcolor={blue!50!black}, linkcolor={red!50!black}]{hyperref}

\begin{document}

\title{The Observation of hyperradiance accompanied by enhanced entanglement in a hybrid optomechanical system}

\author{Zeshan Haider}
 \email{Corresponding author: shani12441@gmail.com}
{\affiliation{National Institute of Lasers and Optronics College, Pakistan Institute of Engineering and Applied Science, Nilore, Islamabad $45650$, Pakistan.}
\author{Muhammad Altaf}
\affiliation{Pakistan Institute of Nuclear Science and Technology, Nilore, Islamabad $45650$, Pakistan}
\author{Tahira Nasreen}
\affiliation{National Institute of Lasers and Optronics College, Pakistan Institute of Engineering and Applied Science, Nilore, Islamabad $45650$, Pakistan.}
 \author{Muhammad Imran}
\affiliation{National Institute of Lasers and Optronics College, Pakistan Institute of Engineering and Applied Science, Nilore, Islamabad $45650$, Pakistan.}
\author{Rameez Ul Islam}
{\affiliation{National Institute of Lasers and Optronics College, Pakistan Institute of Engineering and Applied Science, Nilore, Islamabad $45650$, Pakistan.}
\author{Manzoor Ikram}
{\affiliation{National Institute of Lasers and Optronics College, Pakistan Institute of Engineering and Applied Science, Nilore, Islamabad $45650$, Pakistan.}

\date{\today}

\begin{abstract}
We have theoretically investigated an optomechanical system and presented the scenario of significantly enhanced bipartite photon-phonon entanglement for two qubits coupled to the single mode of the cavity. And results are compared with the one qubit case for reference. The tripartite atoms-photon-phonon interaction is considered as only three-body resonant interaction while the two-body actions are ignored under some potential approximations. Furthermore, we have studied the phenomenon of hyperradiance in which the well-known Dicke superradiant ($N^2$ scaling law) can be surpassed due to the inter-atomic correlations. Jointly, a parameter regime is explored to observe the entanglement of photon-phonon pairs and their hyperradiance simultaneously. As it is important to show that the generation of photons and phonons are antibunched, the equal time second-order correlation function $g^{(2)}(0)$ is characterized as witness. This system can be realized in Circuit Cavity Quantum Electrodynamics (CCQED) in which the direct coupling of the atom and mechanical resonator is possible.
\end{abstract}

\maketitle
\section{introduction}
The study of statistical properties of quantum light and its manipulation in Cavity Quantum Electrodynamics (CQED) \cite{barrett2013simulating,majumdar2012probing,luxmoore2013iii,singh2019quantum} has extensively been the hot area of research in the last few decades. For instance, the photon antibunching leading toward the single and multiple photon sources has been achieved in dipolar microwave cavities \cite{haider2023multiphoton,hou2019interfering,singh2021entanglement}, nanocavities coupled to the quantum dots \cite{liang2020photon,nian2023electrically}, and multimode cavities in thin film carbide photonics \cite{lukin2023two,yang2023inverse} among others. This phenomenon results from the photon blockade effect in which one photon resists the addition of further photons into the system \cite{imamoḡlu1997strongly} and thus can be proven as the potential source of the single photons. It is quite difficult to enumerate the applications of single photons and photon on-demand sources with the recent advancement of quantum information technologies. From quantum communications including Quantum Key Distribution (QKD) \cite{zhang2008secure}, weak force sensing \cite{singh2023enhanced} and teleportation \cite{wang2015quantum} to quantum computing, the applications of these single photon sources are manifold. Therefore, in the optomechanical cavities, the simultaneous emission of photons and phonons and their correlation have been of immense interest \cite{lemonde2013nonlinear,stannigel2012optomechanical}. In such systems, the radiation pressure of the cavity field on one or both movable mirrors can be harnessed to generate the phonons whose correlations with the photons result in various novel effects. On manipulating the optical means, the acoustic excitations (phonons) can effectively be controlled and measured \cite{aspelmeyer2014cavity}. Interestingly, in the recent proposals \cite{shao2023frequency,singh2022tunable}, the frequency conversion of the high-frequency cavity photon into lower-frequency ones plus an interlinked phonon generation is being carried out in an optical cavity where a strongly coupled qubit acts as a mediator. However, in this regard, the complete photon-phonon conversion has already been achieved in a one-dimensional optomechanical lattice via a topologically protected edge channel with a controllable conversion efficiency \cite{cao2021controllable}.

The bipartite and multipartite entanglement has already been proposed in such optomechanical systems with high enough fidelity rate \cite{vitali2007optomechanical, hartmann2008steady,borkje2011proposal,barzanjeh2012reversible, wang2012using, tian2012adiabatic,palomaki2013entangling}. For instance, the bipartite photon-phonon and phonon-phonon entanglement attracted considerable interest with recent advancements of the numerical techniques in optical cavities \cite{akram2012photon, lakhfif2021controlling, xu2019generation, amazioug2020enhancement, araya2023generation}. However, it is quite interesting to entangle the non-gaussian (single) photons and phonons simultaneously under the same parametric regime as proposed in ref. \cite{xu2019generation}. Such optomechanical systems are useful for generating entangled photon-phonon pairs whose rich applications in studies of crystalline solids structures have been duly acknowledged \cite{dekorsy2006coherent,benatti2017generation,grunwald2011raman}.

In the CQED, the cooperative emission from the quantum emitters can be characterized by the collective Dicke states with the Dick superradiance \cite{dicke1954coherence, 
gross1982superradiance} and has been a well-engaged research avenue in the recent past. For example, Pleinert \textit{et al.} \cite{Pleinert2017} proposed that the emission intensity from the emitters surpasses the so-called $N^2$ scaling law of superradiance, the phenomenon they termed as hyperradiance. Later on, the hyperradiance has been achieved for the squeezed light \cite{li2022squeezed} with minimum phase noise in the linear regime having diverse applications in quantum interferometry e.g. to enhance the sensitivity of gravitational wave detectors such as LIGO \cite{aasi2013enhanced} and Geo $600$ \cite{grote2013first}. It is worth mentioning that the position of atoms in the coupled cavity mode becomes quite important to observe this novel effect. Usually, the atoms are asymmetrically coupled in such a way that one is on the crest and the other is on the trough of the cavity mode. However, it has also been achieved with the symmetric atom-field coupling in the recent theoretical investigations \cite{han2021hyperradiance, xu2017hyperradiance}.

Inspired by the above studies, we present the results of an optomechanical system of two identical two-level atoms coupled to the single mode of the optical cavity and a mechanical resonator simultaneously. To the best of our knowledge, the phenomenon of hyperradiance has not been reported in such an optomechanical system so far. We, therefore, theoretically present the tripartite atom-photon-phonon interaction with the exploration of bipartite photon-phonon entanglement accompanied by hyperradiance. The strong antibunched emission witnessed in our system illustrates that the photon-phonon entanglement is potentially non-Gaussian \cite{namiki2012photonic} and thus depends upon the number of atoms and tripartite interaction strength.    
\begin{figure}[htb]
	\includegraphics[width=8.5cm,height=2.5cm]{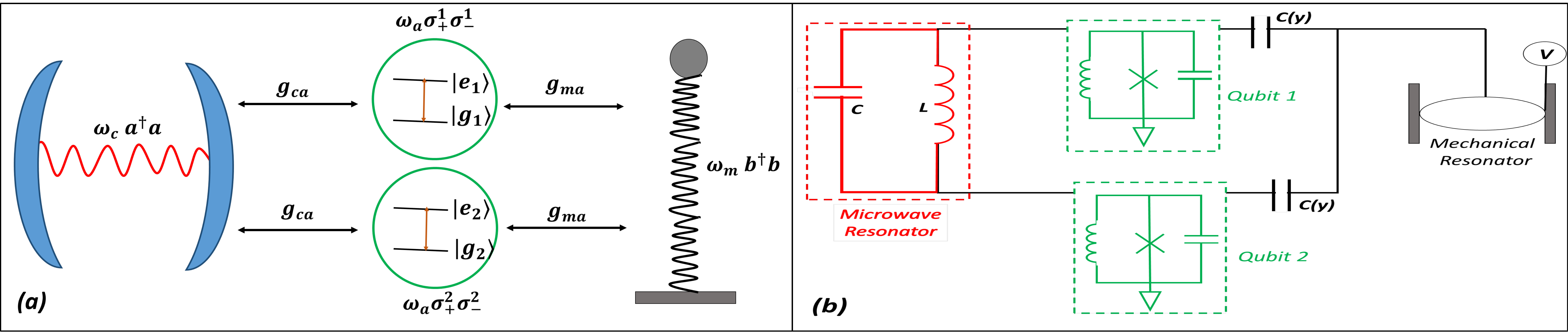}
	\caption{(a) The Schematic diagram of an optomechanical system that is not ideally closed. The two identical two-level atoms are simultaneously interacting with the cavity mode and the mechanical oscillator with the coupling strengths $g_{ca}$ and $g_{ma}$, respectively. (b) The analog circuit cavity QED scheme to (a) for experimental realization. An LC resonator acts as a microwave cavity coupled to the Josephson junction-based qubits (artificial atoms). The qubits are also directly coupled to the AC supply acting as a mechanical oscillator. All three entities, the atoms, cavity, and mechanical oscillator dissipate with the rate of $\kappa$, $\gamma_c$, and $\gamma_m$, respectively. }\label{fig:fig1}
\end{figure}
\section{Model and Dressed State Picture}
We propose a hybrid optomechanical system containing two identical two-level atoms, simultaneously coupled to the single-mode optical cavity and a mechanical resonator with the resonance frequencies of $\omega_c$ and $\omega_m$, respectively as shown in Fig. \ref{fig:fig1} (a). The atoms of the resonant frequency $\omega_a$ are coherently pumped by the classical field of frequency $\omega_p$ that results in the rabi frequency $\Omega$. The coupling of the atoms to the mechanical resonator can be realized in Circuit Cavity QED for the artificial atoms (superconducting Josephson-junction-based qubits) as in ref. \cite{pirkkalainen2015cavity} where the artificial atom is capacitively coupled to the mechanical resonator and is responsible for longitudinal interaction $g_{ma}[\sigma^i_+\sigma^i_-(b+b^\dagger)]$. In the view of the experiment, an analogous circuit Cavity QED diagram of the proposed optomechanical system is shown in Fig. \ref{fig:fig1} (b). Further, both the atoms are also directly coupled to the single mode of the optical cavity, and the weak photon-phonon interaction $g_{mo}[a^\dagger a(b+b^\dagger)]$ can be ignored under the approximation $g_{mo}<<[g_{ma},g_{ca}]$\cite{pirkkalainen2015cavity,heikkila2014enhancing}. It is worth mentioning that the system's evolution is non-unitary due to the significant decay of atoms, photonic, and mechanical modes with the decay rate of $\kappa$, $\gamma_c$ and $\gamma_m$, respectively. For max[$g_{ca},g_{ma}$]$<<$ min[$\omega_0, \omega_m, \omega_c$], the Hamiltonian of the system under rotating wave and dipole approximations reads ($\hbar=1$);
\begin{align}\label{eq:eq1}
H_1&=\omega_c\ a^\dagger a+\omega_m b^\dagger b\\\nonumber
&+\sum_{i=1,2}[\omega_a\sigma^i_+\sigma^i_{-}+g_{ca}(\sigma^i_+a+a^\dagger\sigma^i_-)+g_{ma}[\sigma^i_+\sigma^i_-(b+b^\dagger)]\\\nonumber
&+\Omega(\sigma^i_+e^{-i\omega_pt}+\sigma^i_-e^{i\omega_pt})],
\end{align}with $a$ ($a^\dagger$) and $b$ ($b^\dagger$) being the annihilation (creation) operators of the quantized cavity field and the mechanical mode, respectively. It is assumed that both the atoms feel equal coupling to the quantized cavity field and mechanical resonator having coupling strengths of $g_{ca}$ and $g_{ma}$, respectively with no inter-atomic interaction. The atomic ladder operator for $i^{th}$ atom is labeled as $\sigma^i_\pm$. The atoms are pumped by the classical field of rabi frequency $\Omega$ as shown by the last term in Eq. \ref{eq:eq1}. After applying a unitary displacement transformation $U=\exp[\eta\sum_{i=1,2}(\sigma^i_+\sigma^i_-(b^\dagger-b))]$ with $\eta=g_{ma}/\omega_m$, we obtain the transformed Hamiltonian $H^{'}=UH_1U^\dagger$ as 
\begin{align}\label{eq:eq1bar}
H' &= \omega_c\ a^\dagger a + \omega_m\ b^\dagger b \nonumber \\
&\quad + \sum_{i=1,2} \Big[ (\omega_a - \epsilon)\, \sigma^i_+ \sigma^i_- 
+ g_{ca} \big( a\, \sigma^i_+\, e^{\eta(b^\dagger - b)} 
+ a^\dagger\, \sigma^i_-\, e^{-\eta(b^\dagger - b)} \big) \nonumber \\
&\quad + \Omega \big( \sigma^i_+\, e^{-i\omega_p t}\, e^{\eta(b^\dagger - b)} 
+ \text{H.c.} \big) \Big]
\end{align}
Here, $\epsilon = \frac{g^2_{ma}}{\omega_m}$ introduces the frequency shift under the aforementioned unitary transformation. In view of the experiment, $g_{ma} \ll \omega_m$, i.e., $\eta \ll 1$, and this leads to the approximation $e^{\eta(b^\dagger - b)} \approx 1 + \eta(b^\dagger - b)$. The effective Hamiltonian reads:

\begin{align}\label{eq:eq1bar_approx}
H' &= \omega_c\ a^\dagger a + \omega_m\ b^\dagger b \nonumber \\
&\quad + \sum_{i=1,2} \Big[ (\omega_a - \epsilon)\, \sigma^i_+ \sigma^i_- 
+ g_{ca} \big( a\, \sigma^i_+ + a^\dagger\, \sigma^i_- \big) \nonumber \\
&\quad + J \big( a\, \sigma^i_+ - a^\dagger\, \sigma^i_- \big)(b^\dagger - b) \nonumber \\
&\quad + \Omega \big( \sigma^i_+\, e^{-i\omega_p t} (1 + \eta(b^\dagger - b)) 
+ \text{H.c.} \big) \Big]
\end{align}

with $J=g_{ma}g_{ca}/\omega_m$ denotes the strength of tripartite atom-photon-phonon interaction. 
For $\omega_a-\epsilon\approx\omega_p=\omega_c+\omega_m>>g_{ca}>>\Omega$, the bipartite atom-photon interaction i.e.,  $\sum_{i=1,2}[g_{ca}(a\sigma^i_++a^\dagger\sigma^i_-)]$ and (for large detuning) sideband driving terms $\sum_{i=1,2}\Omega[\sigma^i_+\eta(b^\dagger-b)e^{-i\omega_pt}+H.C]$ can be ignored. The tripartite resonant interaction term i.e., $\sum_{i=1,2}[J(a\sigma^i_+-a^\dagger\sigma^i_-)(b^\dagger-b)]$ can be simplified into $\sum_{i=1,2}[-J(\sigma^i_+ab+\sigma^i_-a^\dagger b^\dagger]$ by neglecting the non-conservative energy terms.
This term depicts the simultaneous generation (absorption) of photon-phonon pair for de-excitation (excitation) of each atom.
In the rotating frame of frequency $\omega_m$ under the unitary transformation matrix
$U(t)=exp[\sum_{i=1,2}i\omega_p\sigma^i_+\sigma^i_{-}t+i(\omega_p-\omega_m)a^\dagger at+i\omega_mb^\dagger bt]$, the Hamiltonian of the system is
\begin{align}\label{eq:eq2}
H&=\Delta\ a^\dagger a\\\nonumber&+\sum_{i=1,2}[\Delta\sigma^i_+\sigma^i_{-}-J(\sigma^i_+ab+\sigma^i_{-}a^\dagger b^\dagger)+\Omega(\sigma^i_++\sigma^i_{-})],
\end{align} where the detuning $\Delta$ is defined as $\Delta=\omega_a-\omega_p=\omega_c-(\omega_p-\omega_m)$ without loss of generality such that $\Delta<<\omega_m$. The above effective Hamiltonian is subjected to the numerical simulations by solving the following Lindblad master equation;
\begin{align}\label{eq:eq3}
\frac{d\rho}{dt}=-\frac{i}{\hbar}\left[H,\rho\right]+{\cal L}_\kappa\rho+{\cal L}_{\gamma_c}\rho+{\cal L}_{\gamma_m}\rho,
\end{align}\label{eq:lmq} under the steady state, and with the help of Quantum toolbox in Python (QuTip) \cite{johansson2012qutip}. In Eq. \ref{eq:eq3}, $\rho$ is the density matrix operator while ${\cal L}_j$ ($j\in$[$\kappa,\gamma_c,\gamma_m$]) is the Liouvillian function incorporating the decays of atoms, cavity and mechanical mode with the decay rate of $\kappa, \gamma_c$, and $\gamma_m$, respectively. In the following, these Liouvillian functions for atoms, cavity, and mechanical mode are defined respectively as
\begin{align}
{\cal L}_\kappa\rho=\kappa\sum_{i=1,2}(2\sigma^i_{-}\rho\sigma^i_{+}-\sigma^i_{+}\sigma^i_{-}\rho-\rho\sigma^i_{+}\sigma^i_{-}),
\end{align}\label{eq:atom}
\begin{align}
{\cal L}_{\gamma_c}\rho=\gamma_c(2a\rho a^\dagger-a^\dagger a\rho-\rho a^\dagger a),
\end{align}\label{eq:cavity}
and
\begin{align}
{\cal L}_{\gamma_m}\rho=\gamma_m(2b\rho b^\dagger-b^\dagger b\rho-\rho b^\dagger b).
\end{align}\label{eq:mechanical}
In the absence of the coherent pumping field i.e., $\Omega=0$, we have calculated the eigenstates of above Hamiltonian for both the case of one and two atoms coupled to the cavity and are presented in the Appendix B along with their basis states (Appendix A). In both cases, the lowest manifold of eigenvalues (one-photon-phonon transitions) is considered and visualized in the dressed state picture shown in Fig. \ref{fig:dressed} (a). For one-atom case, the collective system minimally excites at  $\Delta=\pm J$ and at $\Delta=\pm \sqrt{2}J$ for the two-atom case. In Fig. \ref{fig:dressed} (b), the pathways diagram shows some important transitions of the system. Generally, the notation like in the state $\ket{xycd}$ represents $x,y\in[e,g]$ and $c,d\in[n,m]$ such that $e$ and $g$ are the non degenerate energy levels for two-level atom whereas $n$ and $m$ are the Fock state for cavity and mechenical mode, respectively. It is illustrated that $\Psi^0_0=\ket{gg00}$ is pumped by the classical field via one photon process to the state $\ket{\pm00}=1/\sqrt{2}(\ket{eg00}+\ket{ge00})$ coupled to the state $\ket{gg11}$ through tripartite atom-photon-phonon interaction strength $J$. As the system is open, therefore, the state $\ket{gg11}$ decays to the respective channel shown in Fig. \ref{fig:dressed} (b).

To characterize the statistical properties and antibunching of photons and phonons \cite{amazioug2023strong,singh2021optical}, we compute the equal time second-order correlation functions defined as $g^{(2)}_n(0)$=$\langle a^{\dagger}a^{\dagger}a a\rangle$/$(\langle a^{\dagger} a \rangle)^2$ and $g^{(2)}_m(0)$=$\langle b^{\dagger}b^{\dagger}b b\rangle$/$(\langle b^{\dagger} b \rangle)^2$, respectively. The antibunching (bunching) behaviour can be quantified for $g^{(2)}(0)<1$ ($g^{(2)}(0)>1$). However, the photon-phonon correlation can be calculated through the cross-correlation function i.e., $g^{(2)}_{nm}(0)$=$\langle a^{\dagger}b^{\dagger}b a\rangle$/$(\langle a^{\dagger} a \rangle\langle b^{\dagger} b \rangle)$. The aim of this study is to explore detuning regimes for the entangled and highly correlated photons/phonons, and hence the radiance \cite{Pleinert2017} witness is defined as 
\begin{align}\label{radeq}
R=\frac{\langle a^{\dagger} a \rangle_2-2\langle a^{\dagger} a \rangle_1}{2\langle a^{\dagger} a \rangle_1},
\end{align}\label{eq:rad}
that quantifies the strength of correlated emission from two atoms ($\langle a^{\dagger} a \rangle_2$) w.r.t the single atom coupled twice i.e., $2\langle a^{\dagger} a \rangle_1$. For no correlated emission, $\langle a^{\dagger} a \rangle_2$=$2\langle a^{\dagger} a \rangle_1$ and thus radiance metered to $R=0$. There are three defined regimes based on the parameter $R$ \cite{Pleinert2017} i.e.,$R<0$; subradiance, $0<R<1$; superradiance, and $R>1$; hyperradiance. The phenomenon of superradiance and collective gain in such an optomechanical system has already been studied extensively \cite{kipf2014superradiance, han2019superradiance}. In parallel to the numerical investigation of radiance, we calculate the logarithmic negativity $E_N$, defined as
\begin{align}
E_N=\log_2||\rho^{T_n}_{nm}||_{_1},
\end{align}\label{eq:ENl}where  $||.||_{_1}$ denotes the trace norm and $\rho^{T_n}_{nm}$ is the partial transpose over photonic mode ($T_n$) of reduced density matrix $\rho_{nm}$ associated with the photonic ($n$) and phononic ($m$) modes. It measures ($E_N>0$) the strength of photon-phonon entanglement generated simultaneously through atoms jumping from excited to ground states. It has already been proven in converse that the positivity of the partial transposition of a state is a necessary and sufficient condition for its separability \cite{horodecki1996separability}. The logarithmic negativity then measures the degree to which $\rho^{T_n}$ fails to be positive and can be used as the quantitative version of Peres's criterion \cite{peres1996separability} of partial transpose. In the following, we shall use the logarithmic negativity criteria to diagnose the entanglement as it becomes more practical while numerical investigation of composite systems \cite{vidal2002computable,shapourian2017partial, plenio2005logarithmic}. For the strong tripartite interaction $J$, we present the numerical results demonstrating that $E_N$ can significantly be enhanced with high photon and phonon numbers. 
\begin{figure}[htb!]
	\includegraphics[scale=0.35]{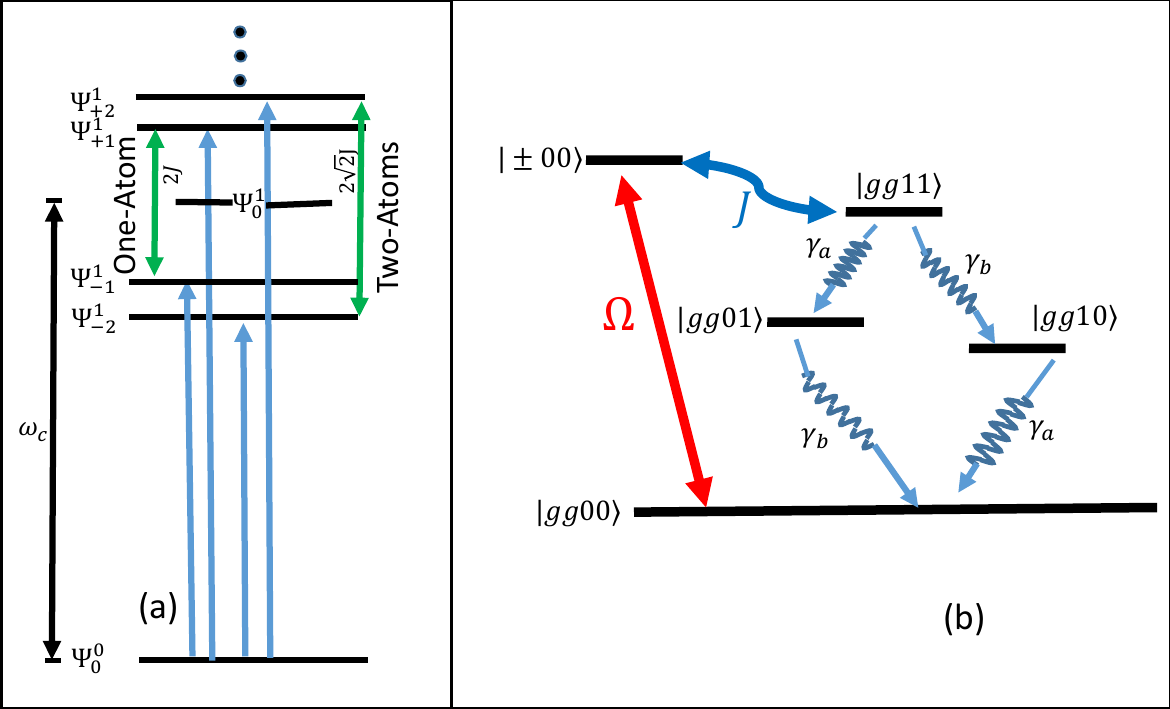}
	\caption{(a) The dressed-state picture up to the first manifold for one-atom and two-atom case under the collective Dick basis. (b) the pathways diagram of some important transitions with $\Omega$ being the pumping strength while $J$ is the strength of tripartite interaction. }\label{fig:dressed}
\end{figure}

% \twocolumngrid

\section{Results and Discussion}\label{sec:TPB}
\begin{figure}[htb]
	\includegraphics[width=\linewidth]{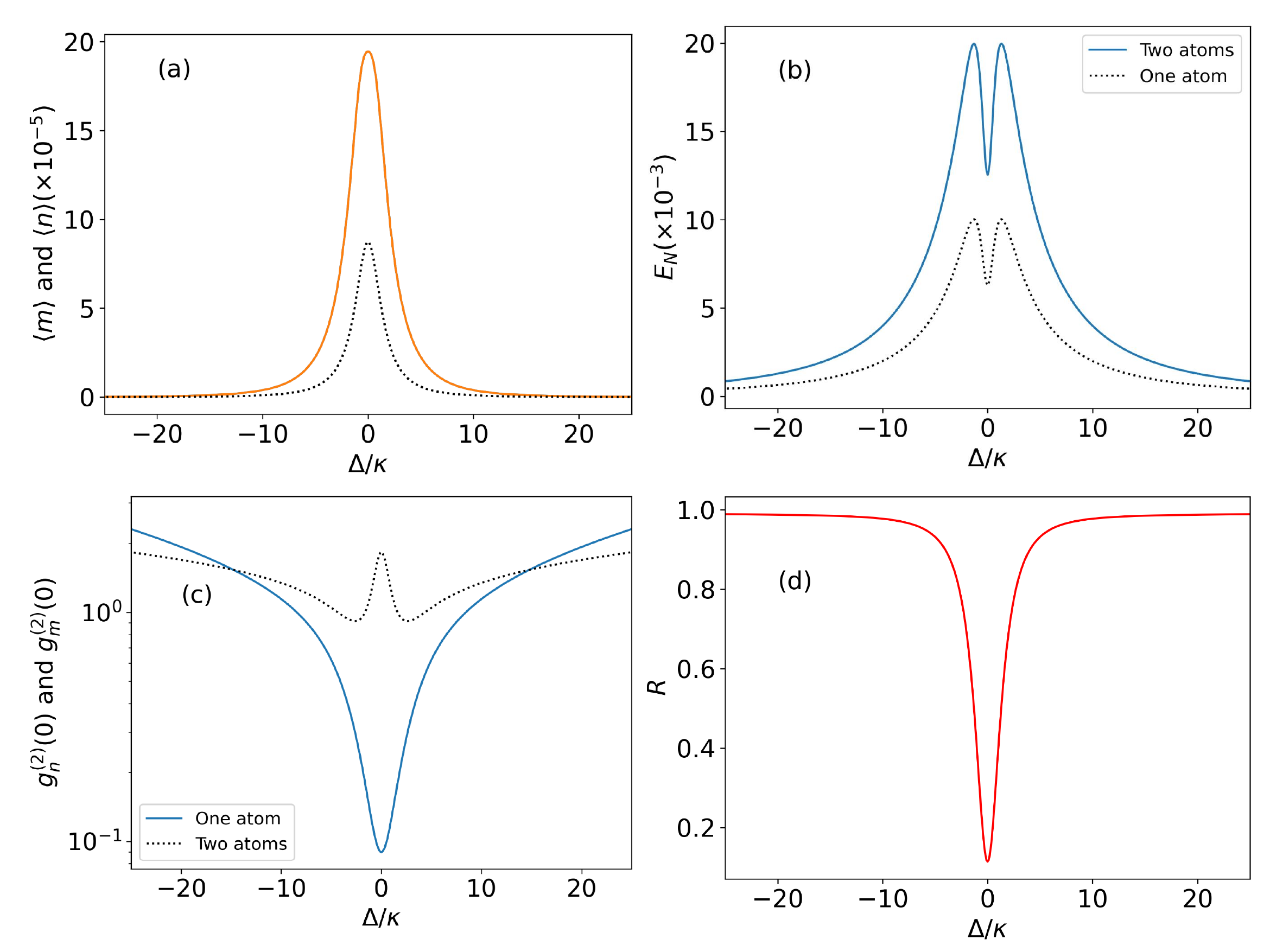}
	\caption{In this Fig, the weak coupling case is being considered i.e., $J=0.1\kappa$. (a) Mean photon (phonon) number [$\langle n \rangle=\langle m \rangle$] is plotted against normalized detuning $\Delta/\kappa$ for one-atom and two-atom case with black dotted and orange solid curves, respectively. In panel (b), the logarithmic negativity $E_N$ while in panel (c) the equal time second-order correlation functions for photons (phonons) [$g^{(2)}_n(0)=g^{(2)}_m(0)$] are plotted as a function of detuning. (d) The radiance witness $R$ against $\Delta/\kappa$ is drawn depicting the strength of correlated emission. We choose $\gamma_c=\gamma_m=10\kappa$ and $\Omega=\kappa$ in all panels.}\label{fig:fig3}
\end{figure}
In this section, we present the results of our numerical simulations. First, we discuss the case of weak coupling of both the atoms with the cavity and mechanical modes (i.e., $J=0.1\kappa$). It is worth noting that atoms are placed at anti-nodes of the cavity mode and thus feel equal atom-field coupling strength. In Fig. \ref{fig:fig3} (a), the mean photon and phonon numbers are plotted against normalized detuning $\Delta/\kappa$ for single and double atoms coupled with the dynamical system. Correspondingly, the logarithmic negativity $E_N$ characterizing the entanglement between photons and phonons is also shown in Fig. \ref{fig:fig3} (b). A dip can be observed at $\Delta=0$ due to the quantum interference between the pathways $\ket{\pm00} \xrightarrow{J}\ket{gg11}$ and $\ket{gg00} \xrightarrow{\eta}\ket{\pm00}\xrightarrow{\eta}\ket{gg00}\xrightarrow{\eta}\ket{\pm00}\xrightarrow{J}\ket{gg11}$. In contrast to the single atom, the significant enhancement of entanglement can be observed in the case of two atoms shown in Fig. \ref{fig:fig3} (b) accompanied by enhanced photon and phonon numbers [see Fig. \ref{fig:fig3} (a)]. As mentioned earlier, excitation (de-excitation) of atoms results in simultaneous absorption (emission) of photons and phonons therefore, $\langle n \rangle$ and $\langle m \rangle$ follow the identical profile as illustrated in Fig. \ref{fig:fig3} (a) for a given span of detuning. Based on the dressed state picture shown in Fig. \ref{fig:dressed} (a), the resonance frequencies for one-atom and two-atom cases are $\Delta/\kappa=\pm J$ and $\Delta/\kappa=\pm\sqrt{2}J$ via channels $\Psi^{(0)}_0 \rightarrow \Psi^{(1)}_{\pm 1} $ and $\Psi^{(0)}_0 \rightarrow \Psi^{(1)}_{\pm 2} $, respectively. At these transitions, the non-negative value of $E_N$ shown in Fig. \ref{fig:fig3} (b) depicts that simultaneously generated photons and phonons are entangled for both cases of single and double atoms. However, for the case of two atoms, photon-phonon entanglement significantly increases and becomes double in comparison to the case of single-atom. The purity of emitted photons and phonons is characterized by their corresponding equal time second-order correlation functions $g^{(2)}_n(0)$ and $g^{(2)}_m(0)$ plotted in Fig. \ref{fig:fig3} (c). It is evident from the second-order correlation functions in Fig. \ref{fig:fig3} (c) that the entangled photons and phonons are antibunched [$g^{(2)}_n(0)=g^{(2)}_m(0)<1$] for one-atom and bunched [$g^{(2)}_n(0)=g^{(2)}_m(0)>1$] for two-atom system. To achieve the antibunched emission for the two-atom case as well, one needs to invoke strong coupling regime as discussed below. We also characterize the strength of photonic emission in Fig. \ref{fig:fig3} (d) by numerically computing the radiance as described in Eq. \ref{radeq}. Since the atom-field coupling is weak ($J=0.1$), the one-photon manifold excites at the nearly same detuning ($\Delta/\kappa=\pm\sqrt{2}J\approx \pm J$ for small $J$) in both the cases of one-atom and two-atom systems. This results in the sub-radiant ($R<1$) with photons entangled to the phonons, as shown in Fig. \ref{fig:fig3} (d). The strength of this correlated emission of photons can be enhanced to hyperradiance regime by using the strong atom-field coupling as presented below.
\begin{figure}[htb]
	\includegraphics[width=\linewidth]{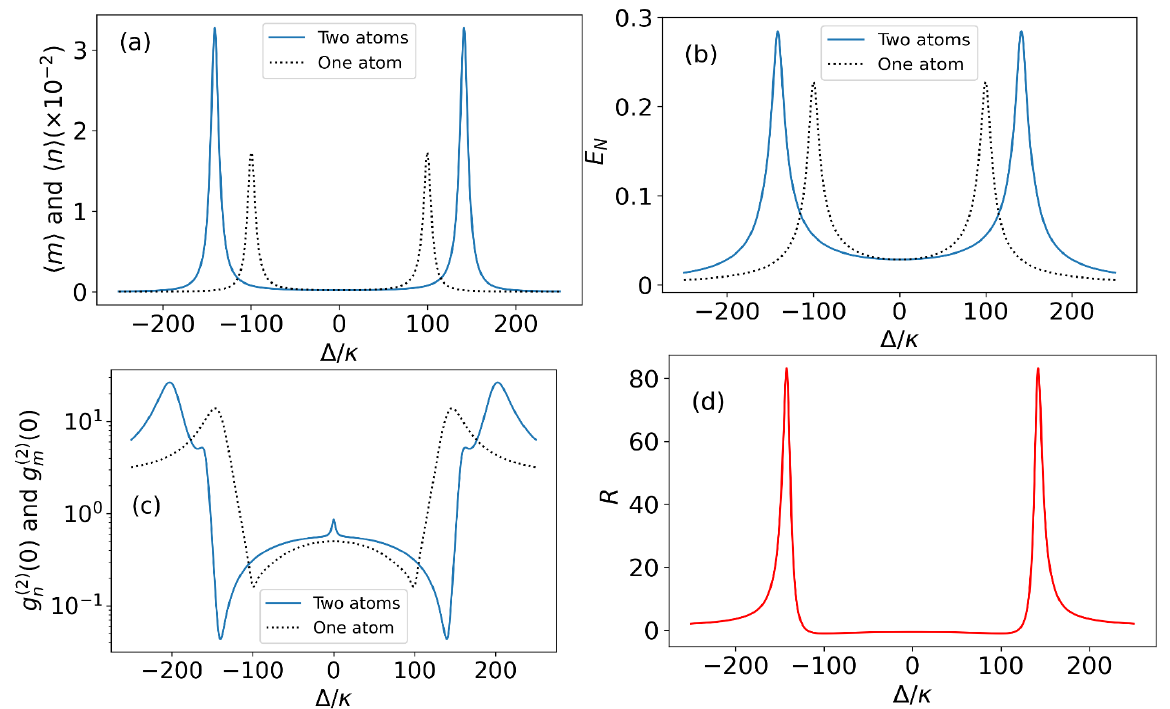}
	\caption{For the strong tripartite interaction strength $J=100\kappa$, mean photon (phonon) number [$\langle n \rangle=\langle m \rangle$], logarithmic negativity $E_N$, the equal time second-order correlation functions for photons (phonons) [$g^{(2)}_n(0)=g^{(2)}_m(0)$], and Radiance $R$ is plotted against normalized detuning $\Delta/\kappa$ in panel (a), (b), (c), and (d), respectively. The rest of parameters are same as used in Fig. in \ref{fig:fig3}.   }\label{fig:fig4}
\end{figure}

In the realm of strong coupling ($J=100\kappa$), Fig. \ref{fig:fig4} shows the dynamical behavior of the system in correspondence to Fig. \ref{fig:fig3}. In contrast with the weak-coupling case, the resonant frequencies of the collective system become more spaced i.e., at $\Delta/\kappa=\pm J$ for one-atom and at $\Delta/\kappa=\pm \sqrt{2}J$ for two-atoms case. In both the weak ($J<\kappa$) and strong ($J>\kappa$) coupling regimes, the photon-phonon entanglement has significantly been enhanced for the case of two-atoms along with their enhanced photon and phonon numbers as sketched in Fig. \ref{fig:fig4} (a) and (b). It is due to the correlated emission of multiple coupled atoms and is not available in uncorrelated one-atom case. The purity of single photonic and phononic emissions can be noted from equal-time second-order correlation functions in Fig \ref{fig:fig4} (c). At the transition frequencies, [$g^{(2)}_n(0)$ and  $g^{(2)}_m(0)$]$_{two-atoms}<$[$g^{(2)}_n(0)$ and  $g^{(2)}_m(0)$]$_{one-atom}$ depicting that in contrast to the one-atom, emission is more antibunched and non-classical in the case of two-atoms. Furthermore, these correlation functions also provide an evidence of single as well as discrete emission by the collective system. Interestingly, in this case of strong coupling ($J>\kappa$) the system undergoes the phenomenon of hyperradiance ($R>1$) at the frequencies $\Delta/\kappa=\pm\sqrt{2}J$ as depicted in Fig. \ref{fig:fig4} (d). This also quantifies the strength of correlated emission by multiple atoms in comparison with the uncorrelated one as followed by Eq. \ref{radeq}. One cannot observe the hyperradiance in the weak coupling ($J<\kappa$) regime because photonic and phononic emission for both the one-atom and two-atoms cases appears at roughly identical frequencies with the identical emission spectrum as shown in Fig. \ref{fig:fig3} (a). The energy difference between the states $\Psi^{(1)}_{\pm1}$ and  $\Psi^{(1)}_{\pm2}$ is $\delta/\kappa=(\sqrt{2}-1)J$ and can be tuned non-linearly through $J$ as shown in the dressed state picture shown in Fig. \ref{fig:dressed} (a). Therefore, resonant frequencies for one-atom and two-atoms cases can be tuned and shifted with the gap of $41.4\kappa$ for $J=100\kappa$. This energy gap results in the observation of hyperradiance.

\begin{figure}[htb]
	\includegraphics[width=\linewidth]{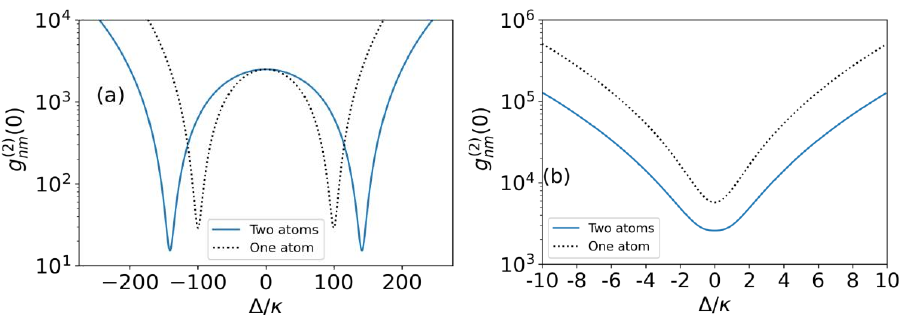}
	\caption{The second-order cross-correlation function $g^{(2)}_{nm}(0)$ between photons and phonons is plotted against normalized detuning for both strong ($J=100\kappa$) and weak ($J=0.1\kappa$) coupling in panel (a) and (b), respectively. the rest of the parameters are the same as in \ref{fig:fig3}.}\label{fig:cross}
\end{figure}

Quite opposite to the pure states, the quantum correlation and entanglement are relatively detached in the case of the mixed photon-phonon states \cite{vesperini2023entanglement, xi2015quantum, adesso2016measures,singh2014quantum}. Therefore, to explore further, second order cross-correlation function $g^{(2)}_{nm}(0)$ between photons and phonon is plotted against normalized detuning for strong and weak coupling regimes in Fig. \ref{fig:cross} (a) and (b), respectively. As the photon-phonon cross-correlation function get suppressed with enhanced photon/phonon number, and entanglement at resonance therefore, in the strong coupling regime in Fig. \ref{fig:cross} (a), photons and phonons are maximally uncorrelated at the $\Delta=\pm\sqrt{2}J$ and hence, maximally entangled as shown in Fig. \ref{fig:fig4} (b). Similarly, in the case of two weakly coupled atoms, the photon-phonon correlation significantly reduces at resonance as shown by the blue solid curve in Fig. \ref{fig:cross} (b). This emphasizes how one can observe the noticeable enhancement of photon-phonon entanglement generated simultaneously at the mentioned detunings in the presented scheme comprized of the two atoms.
\begin{figure}[htb]
     \includegraphics[width=\linewidth]{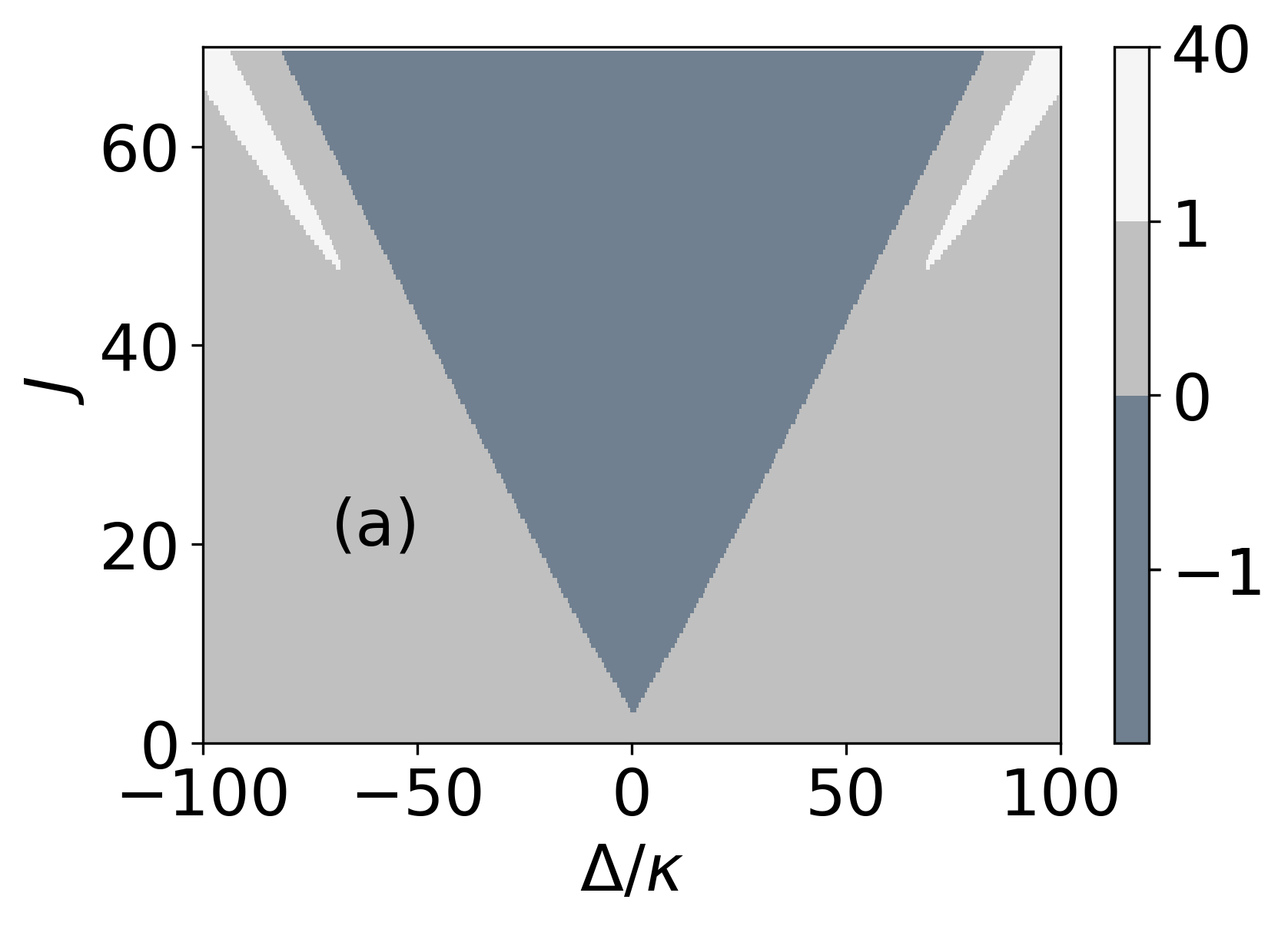}
	\includegraphics[width=\linewidth]{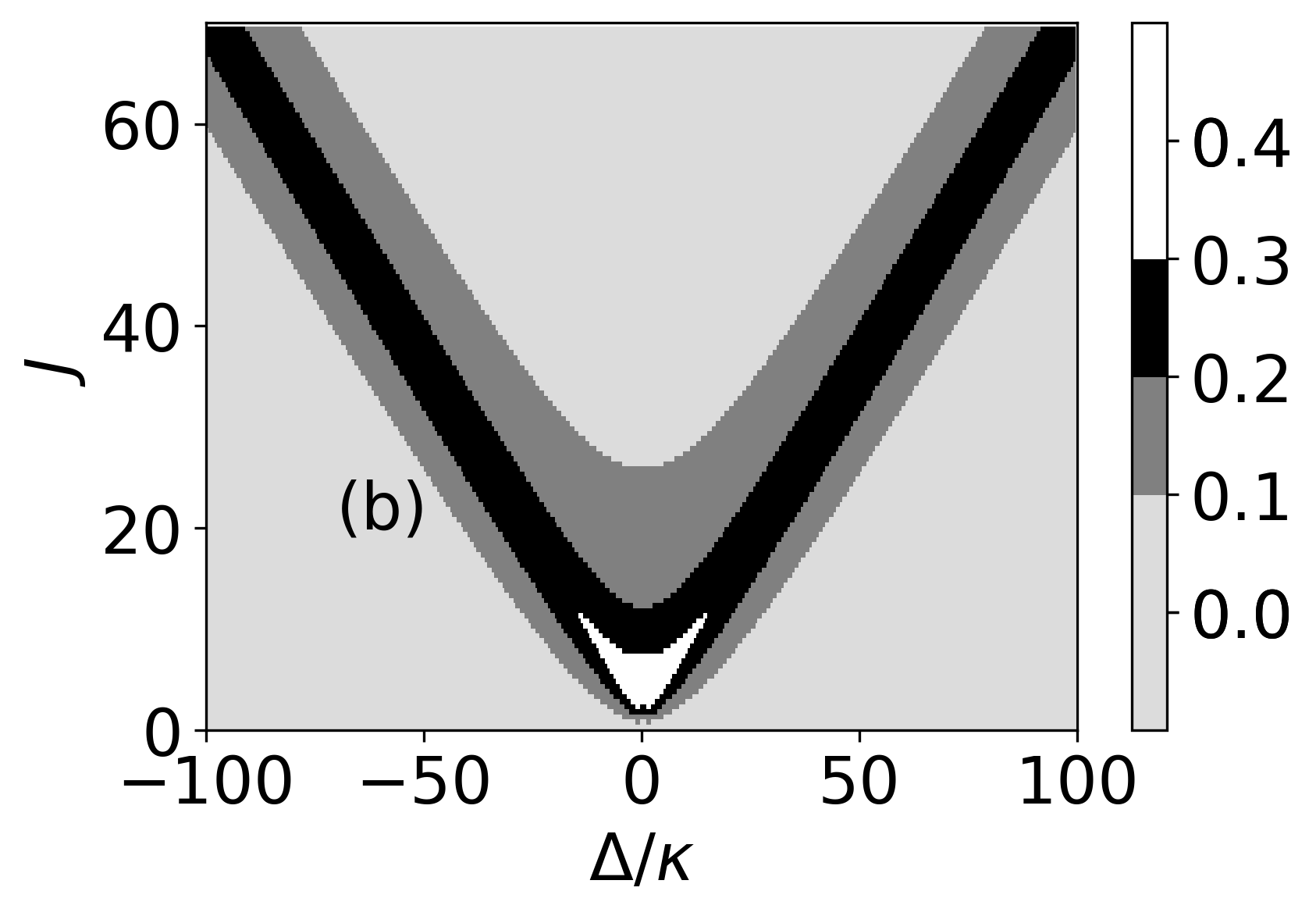}
	\caption{The radiance $R$ (a) and corresponding logarithmic negativity $E_N$ (b) is plotted as a function of coupling strength $J$ and normalized detuning $\Delta/\kappa$ with the remaining parameters as used in Fig. \ref{fig:fig3}. }\label{fig:fig6}
\end{figure}

Next, we investigate our numerical simulations concerning radiance which clearly demonstrate that the presented scheme is quite favorable for achieving the novel phenomenon of hyperradiance with the entangled photons. In Fig. \ref{fig:fig6} (a), radiance is plotted against coupling strength $J$ and the normalized detuning. The darkest region shows the phenomenon of sub-radiance ($R<0$) accompanied by the gray region associated with the superradiance ($0<R<1$). The most correlated photonic/phononic emission is achieved at white clefts symmetrically in detuning space and is termed as hyperradiance ($R>1$). It is worth noting that hyperradiant photons and phonons are also strongly mutually entangled. As an evidence, corresponding to the white hyperradiant regions in Fig. \ref{fig:fig6}(a), there is a stronger regime (black) of photon-phonon entanglement shown in Fig. \ref{fig:fig6} (b). However, the strongest entangled correlations are observed in the white region of Fig. \ref{fig:fig6} (b) where coupling strength $J$ becomes comparable to the decay rates of photonic (phononic) mode i.e., $J\approx \gamma_c$ =$\gamma_m$. The white region is not found favourable for achieving the excessively correlated emission such as hyperradiance as the one-atom and two-atoms transitions occur with the narrow detuning gap of $\delta/\kappa=(\sqrt{2}-1)J$. As the $J$ increases along the vertical axis of Fig. \ref{fig:fig6} (a), the mentioned detuning gap becomes significant and enhances the radiance drastically. In Fig. \ref{fig:mpn}, we further explore how the presented scheme stimulates the hyperradiant emission accompanied by stronger photon-phonon entanglement. As mentioned earlier, one-atom and two-atom transitions take place at the same detuning in the vicinity of resonance ($-10\kappa\leq\Delta\leq10\kappa$) and move further apart as the coupling strength $J$ gets stronger, and this is illustrated in Fig. \ref{fig:mpn}. On comparison the one-atom and two-atom excitation spectrum in Fig. \ref{fig:mpn} (a) and (b) respectively, it might be noted that the opening of the V-shaped spectrum is relatively broader in the case of two-atom system for strong coupling shown Fig. \ref{fig:mpn} (b). It provokes the system to manifest hyperradiance at a strong coupling regime. From the experimental point of view, we believe that such a dependence of correlated emission on atom-field coupling may attract significant attention while tuning the emission from subradiance to superradiance and hyperradiance through the manipulation of the $J$ value.
\begin{figure}[htb]
	\includegraphics[scale=0.29]{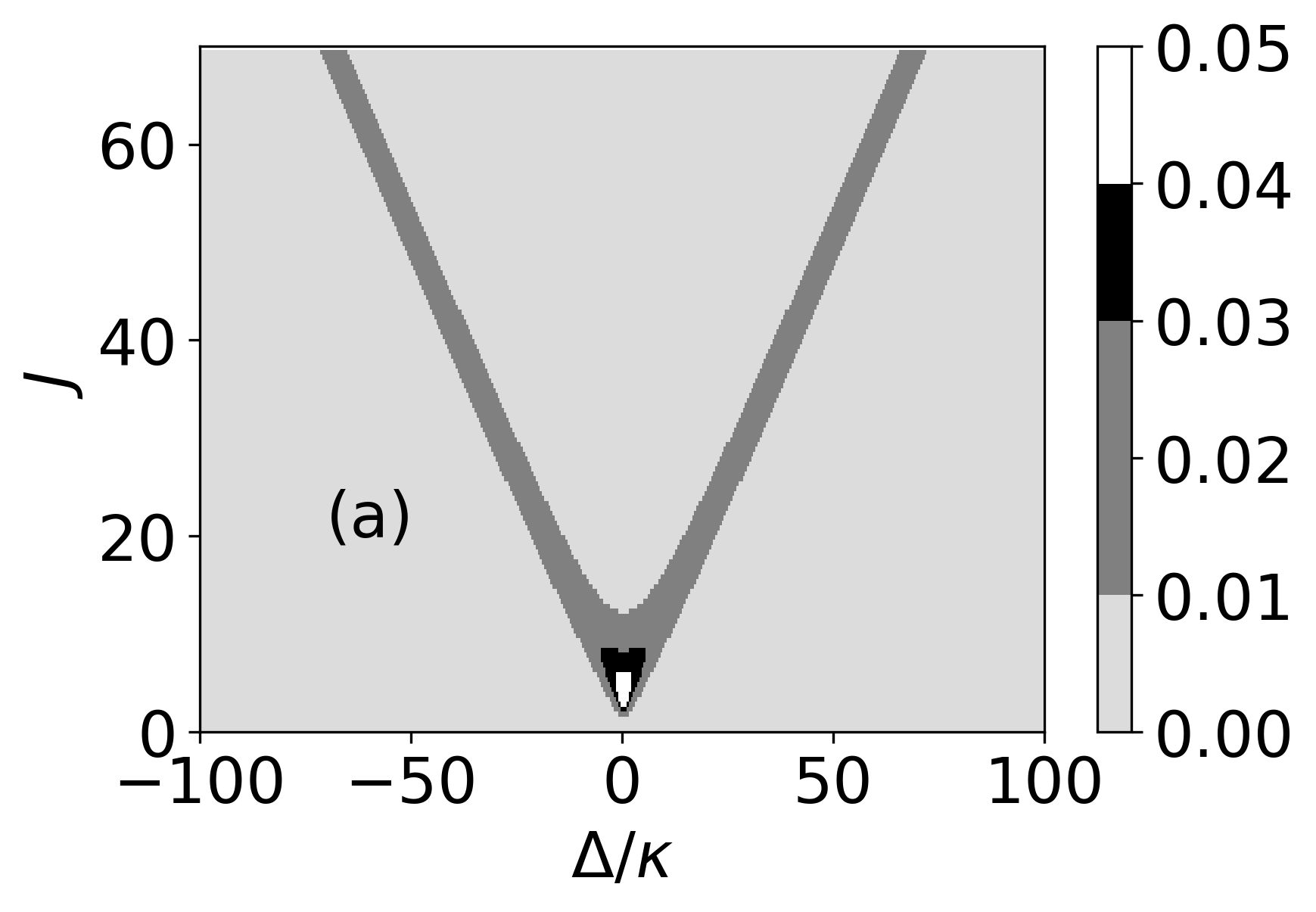}
    \includegraphics[scale=0.29]{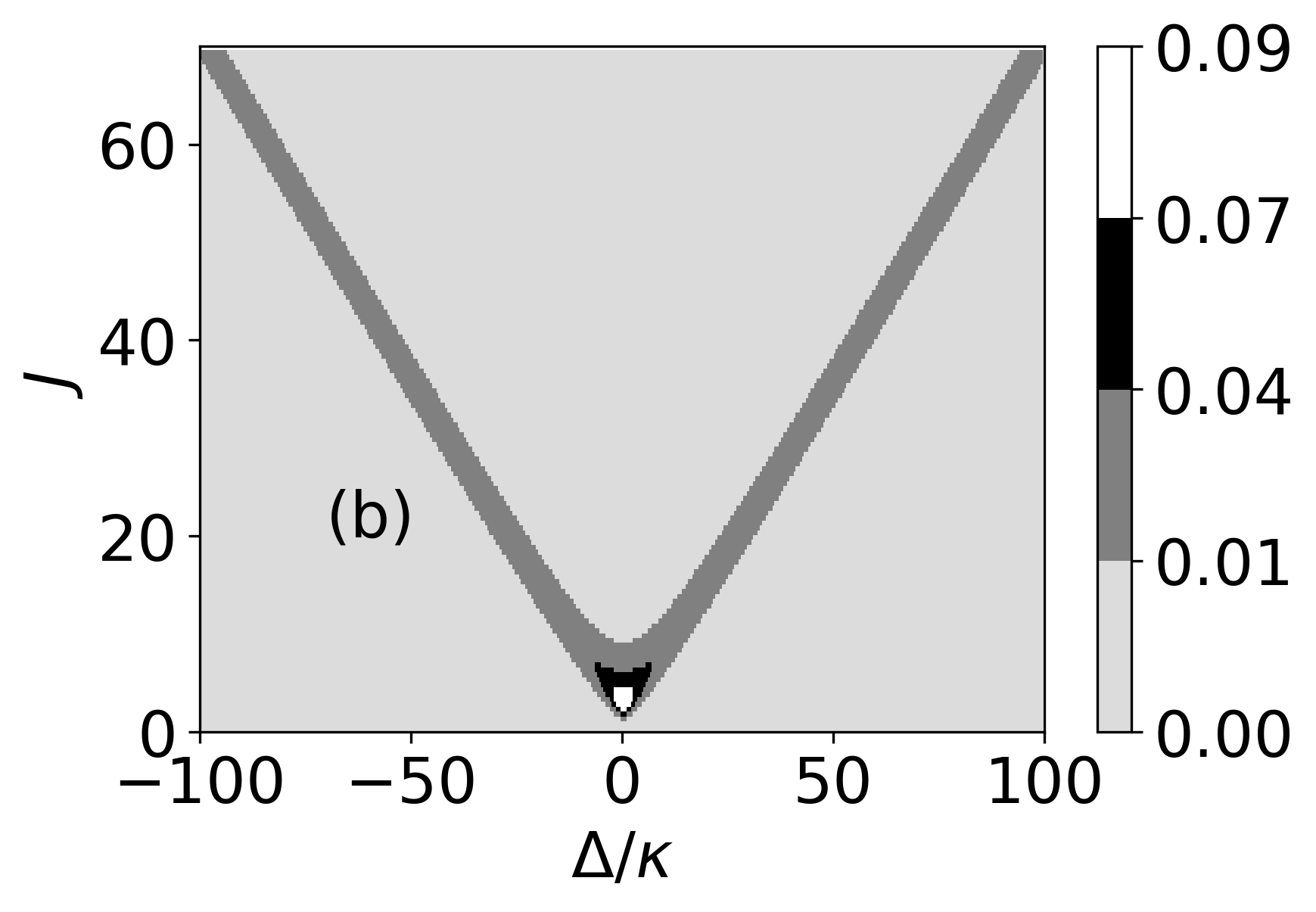}
	\caption{Mean photon and phonon number [$\langle n \rangle=\langle m \rangle$] for one-atom in panel (a) and two-atoms in panel (b) as a function of coupling strength $J$ and detuning $\Delta/\kappa$ is plotted. }\label{fig:mpn}
\end{figure}
\section{Conclusion}\label{sec:conc}
It is concluded that photon-phonon entanglement in the proposed system can be enhanced by coupling the two qubits with the single mode of the cavity rather than one qubit in weak and strong coupling ($J$) regimes. In both one-qubit and two-qubit cases, the strong coupling strength $J$ causes the larger vacuum Rabi splitting which provides a chance to observe the hyperradiance with enhanced mean photon/phonon number. The non-classical emission of photons and phonons ($g^{(2)}(0)<1$) shows that their entanglement is purely non-gaussian. We believe that the proposed system is quite feasible for experimental realization in the microwave regime as presented in ref. \cite{pirkkalainen2015cavity}. The applications of pure and maximally entangled photon-phonon pairs are manifold such as in hybrid quantum networks to connect the optical communication media to quantum memories \cite{mol2023quantum}, quantum teleportation \cite{PhysRevLett.82.1056} among others. 
\section{Data Availability Statement}
As all data has been presented in the main text graphically, therefore this manuscript has no associated data information.
\begin{widetext}
\appendix
\section{Definition of Basis Sates for one-atom and two-atoms system}
The collective basis states in $1$-photon manifold for one-atom system are $\ket{g00}$, $\ket{g10}$, $\ket{g01}$, and $\ket{00}_{\pm}$. However, for the two-atoms system, these basis states are defined as $\ket{gg00}$, $\ket{gg10}$, $\ket{gg01}$, $\ket{gg11}$, and $\ket{\pm00}$
The entangled states are defined as:
\begin{equation}
\ket{00}_{\pm}=\frac{\ket{g11}\pm\ket{e00}}{\sqrt{2}},
\end{equation}
and
\begin{equation}
\ket{\pm00}=\frac{\ket{ge00}\pm\ket{eg00}}{\sqrt{2}}
\end{equation}
\section{Eigenvalues and Eigensates of one-atom and two-atom systems}
On diagonalizing the Hamiltonian presented in the main text for both the case of one-atom and two-atom system, the eigenvalues and their corresponding eigenfunction are presented in the table \ref{tb:tabel1} and table \ref{tb:tabel2}, respectively. The dressed state diagram in Fig. \ref{fig:dressed} (a) is constructed based on these eigenstates.
\begin{table}[ht]
    \centering
    \begin{tabular}{|c|c|}
    \hline
        Eigenvalues & Eigenstates\\
    \hline
       $\lambda_0=\omega_c$ & $\Phi^{(1)}_0=\ket{g00}$, $\Phi^{(2)}_0=\ket{g01}$, and $\Phi^{(3)}_0=\ket{g10}$\\
    \hline
       $\lambda^{(1)}_{\pm1}=\omega_c\pm J$ & $\Psi^{(1)}_{\pm1}=\frac{\ket{g11}\pm\ket{e00}}{\sqrt{2}}$\\
    \hline
    \end{tabular}
    \caption{One-Atom System}
    \label{tb:tabel1}
\end{table}
\begin{table}[ht]
    \centering
    \begin{tabular}{|c|c|}
    \hline
        Eigenvalues & Eigenstates\\
    \hline
       $\lambda_0=\omega_c$ & $\Psi^{(1)}_0=\ket{gg00}$, $\Psi^{(2)}_0=\ket{gg01}$, $\Psi^{(3)}_0=\ket{gg10}$, and $\Psi^{(4)}_0=\ket{-00}$\\
    \hline
       $\lambda^{(1)}_{\pm2}=\omega_c\pm \sqrt{2}J$ & $\Psi^{(1)}_{\pm2}=\frac{\ket{gg11}\pm\ket{+00}}{\sqrt{2}}$\\
    \hline
    \end{tabular}
    \caption{Two-Atom System}
    \label{tb:tabel2}
\end{table}
\end{widetext}

\ifdefined\ispreprint
  \bibliographystyle{apsrev4-2}
  \bibliography{ref}
\else

\fi

\end{document}